\documentclass[fleqn,usenatbib]{mnras}

\usepackage{graphicx}
\usepackage{txfonts}

\usepackage[table]{xcolor}

\newcommand{\rp}{G_\mathrm{RP}}
\newcommand{\vsini}{v\sin i}

\newcommand{\halpha}{H_\mathrm{\alpha}}
\newcommand{\pnd}{\mathrm{P}_\mathrm{ND}}

   \title[UVdim stars in Galactic OCs]{Hunting for UVdim stars in Galactic Open clusters. Clues from ultraviolet photometry}

   \author[G. Cordoni et al.]{G. Cordoni,$^{1}$\thanks{E-mail: giacomo.cordoni@anu.edu.au},
          A.~P. Milone$^{2, 3}$,
          L.Casagrande$^{1}$,
          L. Venuti$^{4, 5}$,
          E.~P. Lagioia$^{6}$,
          F. Muratore$^{2}$,
         \newauthor
          A.~F. Marino$^{3,7}$,
          G.~S. Da Costa$^{1}$,
          F. Dell'Agli$^{8}$,
          F. D'Antona$^{8}$
              \\
    $^{1}$ Research School of Astronomy and Astrophysics, The Australian National University, Canberra, ACT 2611, Australia \\
    $^{2}$Dipartimento di Fisica e Astronomia''Galileo Galilei`` - Univ. di Padova, Vicolo dell'Osservatorio 3, Padova, IT-35122 \\
    $^{3}$Istituto Nazionale di Astrofisica - Osservatorio Astronomico di Padova, Vicolo dell'Osservatorio 5, Padova, IT-35122 \\
    $^{4}$SETI Institute, 339 Bernardo Ave., Suite 200, Mountain View, CA 94043, USA \\
    $^{5}$Visiting Fellow, School of Physics, UNSW Science, Kensington, NSW 2052, Australia \\
    $^{6}$South-Western Institute for Astronomy Research Yunnan University, Kunming, 650500, P.R. China \\
    $^{7}$Istituto Nazionale di Astrofisica - Osservatorio Astrofisico di Arcetri, Largo Enrico Fermi, 5, Firenze, IT-50125 \\
    $^{8}$Istituto Nazionale di Astrofisica, Osservatorio Astronomico di Roma, Via Frascati 33, 00077 Monte Porzio Catone, Italy
    }

\date{Accepted XXX. Received YYY; in original form ZZZ}

\pubyear{2024}

\begin{document}
\label{firstpage}
\maketitle

\begin{abstract}
    
        Split main-sequences (MSs) and extended main-sequence turn-offs (eMSTOs) have been observed in nearly all Magellanic Clouds clusters younger than 2\,Gyr. More recently, Hubble Space Telescope (HST) ultraviolet photometry uncovered a puzzling new population of UV-absorbed stars, dubbed ``UVdim'', in five Magellanic Clouds clusters aged between 40 and 200\,Myr, as well as in one 1.5\,Gyr-old cluster. These UVdim stars predominantly lie on the blue MS, which is composed of slow rotators, and their distinct UV properties are believed to stem from dusty circumstellar disks.

        Although eMSTOs are common in both Magellanic Clouds and Galactic open clusters (OCs) of comparable ages, UVdim stars have not yet been investigated in Galactic OCs. In this work, we fill that gap by combining  \textit{Swift}/UVOT, SkyMapper, and Gaia photometry to extend the search for UVdim stars to 35 Galactic OCs younger than 2\,Gyr. 
        
        By constructing colour–colour diagrams analogous to those employed with HST WFC3/UVIS, we find no evidence of UVdim-like stars in most Galactic open clusters and identify possible UVdim candidates in only five systems.The rarity of UVdim stars in young OCs suggests a potential difference between Magellanic Cloud clusters and their Milky Way counterparts, although the underlying reason remains unclear.

\end{abstract}

\begin{keywords}
stars: Hertzsprung-Russell and colour-magnitude diagrams -- evolution; Galaxy: open clusters and associations: general -- open clusters and associations: individual
\end{keywords}
%

\section{Introduction}\label{sec:intro}
Over the past two decades, our understanding of young star clusters has dramatically changed. High-precision optical and near-ultraviolet (NUV) photometry from the Hubble Space Telescope (HST) uncovered unexpected features in the colour-magnitude diagrams (CMDs) of young star clusters in the Large and Small Magellanic Clouds \citep[SMC, LMC][and series]{bastian2009, milone2009}. Clusters aged between approximately between 700 Myr and 2 Gyr have narrow main sequences but extended main-sequence turn-offs \citep[eMSTOs][]{mackey2007, bastian2009, milone2009, goudfrooij2014, milone2023a}, which cannot be explained by unresolved binaries or photometric errors.
Furthermore, clusters younger than about 700 Myr show split or broad main sequences \citep[MSs,][]{milone2016, milone2023a}, hosting  a blue and a red MS (bMS, rMS).

Several studies have explored the origin of the split-MS \citep[see e.g.][]{milone2009, milone2018, bastian2009, bastian2015, goudfrooij2014, goudfrooij2017, niederhofer2015, dantona2017, ettorre2025}, leading to a widespread interpretation that split-MSs and eMSTOs result from the presence of stars with different rotation rates. These rotational differences are responsible for both the split or broadening of the MS and the colour spread in the turn-off region \citep{georgy2013}. This interpretation was later confirmed by spectroscopic studies, which directly measured projected stellar rotation ($\vsini$) in MS and turn-off stars \citep[see e.g.][]{dupree2017, marino2018a, marino2018b, kamann2020, kamann2023}, finding that slow and fast rotating stars are predominantly located on the bMS and rMS, respectively.

More recently, data from the Gaia mission \citep{gaiadr2, gaiadr3} demonstrated that eMSTOs and split main sequences are also found in young Galactic Open clusters (OCs), and not just in Magellanic Clouds clusters \citep[see e.g., ][and references therein]{marino2018b, bastian2018, cordoni2018, sun2019, li2024, bu2024, cordoni2024}. However, while nearly all MCs clusters younger than about 700 Myr show split-MS \citep[see e.g., Fig. 13 of][]{milone2023a}, only a few OCs do (NGC\,2287, NGC\,3532, and possibly NGC\,2548 \citep[see e.g.][]{sun2019, cordoni2024}. 
This raises questions about the origin of the split-MS and what causes the differences between OCs and MCs clusters.

While rotation is widely accepted as the cause of split MSs, the origin of the bimodal rotation distribution, and overall rotational differences, driving this phenomenon remain unclear and debated. Several explanations have been proposed: \textit{i)} all stars initially form as fast rotators but slow down due to tidal interactions \citep[see, e.g.,][for details and tests]{dantona2015, sun2019, bu2024, muratore2024}; \textit{ii)} a bimodal rotation distribution arising from variations in pre MS circumstellar disks lifetimes and the disk-locking mechanism \citep{bastian2020, bu2024b}; and \textit{iii)} stellar mergers and diverse evolutionary paths contributing to the observed eMSTOs and split-MSs \citep{wang2022}.

\subsection{UVdim stars in Magellanic Clouds clusters}

Recent studies have identified a new feature in young and intermediate-age MC clusters. Using NUV imaging with HST F225W/F275W filters, \citet{milone2023a, milone2023b} discovered a population of stars exhibiting strong UV absorption, dubbed UVdim stars. These UVdim stars were first detected in the 1.5\,Gyr-old LMC cluster NGC\,1783. Their photometric properties suggest the presence of edge-on dusty rings, likely resulting from excretion discs expelled by fast-rotating stars during mass-loss events\citep{dantona2023}.

More recently \citet{milone2023b} and \citet{martocchia2023} extended the analysis to other Magellanic Cloud clusters younger than 200\,Myr, and older than 2\,Gyr respectively. By exploiting a two-colour diagram ($m_\mathrm{F225W}-m_\mathrm{F336W}$ vs. $m_\mathrm{F336W}-m_\mathrm{F814W}$), \citet{milone2023b} identified UVdim stars within these younger clusters. UVdim stars in these young clusters predominantly lie on the blue main sequence\footnote{The blue main sequence is traditionally defined as the bluer sequence in the $m_\mathrm{F814W}$ vs. $m_\mathrm{F336W} - m_\mathrm{F814W}$ CMD.}, which is primarily composed of slower rotators. 

It is noteworthy that, although photometric studies suggest negligible rotation rates for blue-MS stars, spectroscopic observations indicate a more complex scenario. Specifically, blue-MS stars exhibit lower rotational velocities compared to red-MS stars, yet span a broad range of rotation speeds \citep[e.g.,][]{kamann2023,bodensteiner2023}. Supporting the low-rotation nature of UVdim stars, \citet{leanza2025} recently analysed MUSE data from the 1.5\,Gyr cluster NGC\,1783 in the Large Magellanic Cloud, finding UVdim stars consistently rotate more slowly than the majority of main-sequence stars.

.
On the other hand, \citet{kamann2023} analysed MUSE spectroscopy of the LMC cluster NGC\,1850, suggesting that UVdim stars could be Be stars—rapidly rotating B-type stars observed edge-on through circumstellar decretion disks. Additionally, UVdim stars appear to vanish in clusters older than 2\,Gyr, a behaviour analogous to the disappearance of the eMSTO phenomenon \citep{martocchia2023}.

The nature of UVdim stars is therefore still highly debated, and there is no consensus about their rotating or non-rotating origin.

Together, these recent developments suggest a clear connection between split MSs/eMSTOs, stellar rotation, and circumstellar disks. However, the exact nature of this relationship, and the origin of UVdim stars, remains unclear. Further observational and theoretical efforts are needed to better understand the formation of UVdim stars and their role in shaping cluster CMDs.

At the same time, key open questions remain regarding the broader context of young star clusters. While the eMSTO is commonly observed in both Magellanic Cloud clusters and Galactic OCs, split MSs appear to be rarer among OCs. What accounts for this difference? And are UVdim stars present in young Galactic OCs as well? Addressing these questions is crucial to disentangling the physical mechanisms underlying multiple populations in young Galactic and Magellanic Clouds star clusters and their connection with the environment.

In this study, we extend the study of UVdim stars to young Galactic OCs by combining ultraviolet photometry from the \textit{Swift}/Ultraviolet-Optical Telescope Stars (UVOT) Stars Survey \citep{siegel2014, siegel2019}, SkyMapper Southern Sky Survey Data Release 4 \citep[DR4]{onken2024}\footnote{DR4 release: \url{10.25914/5M47-S621}}, and Gaia DR3 \citep{gaiadr3}. This paper is organised as follows: Sec.~\ref{sec:data} describes the data and selection criteria, Sec.~\ref{sec:results} presents our results, and Sec.~\ref{sec:discussion} discusses our findings and conclusions.

\section{Data and data selection} \label{sec:data}

To investigate the presence of UVdim stars in young Galactic OCs and compare them with Magellanic Cloud clusters observed by HST, we used ultraviolet NUV photometry from the Swift Open Cluster Survey, $u$-band photometry from the SkyMapper DR4, and $\rp$ photometry from Gaia DR3. We refer to the references listed in Sec.~\ref{sec:intro} for details on the surveys and data. 

\begin{figure}
    \centering
    \includegraphics[width=0.4\textwidth]{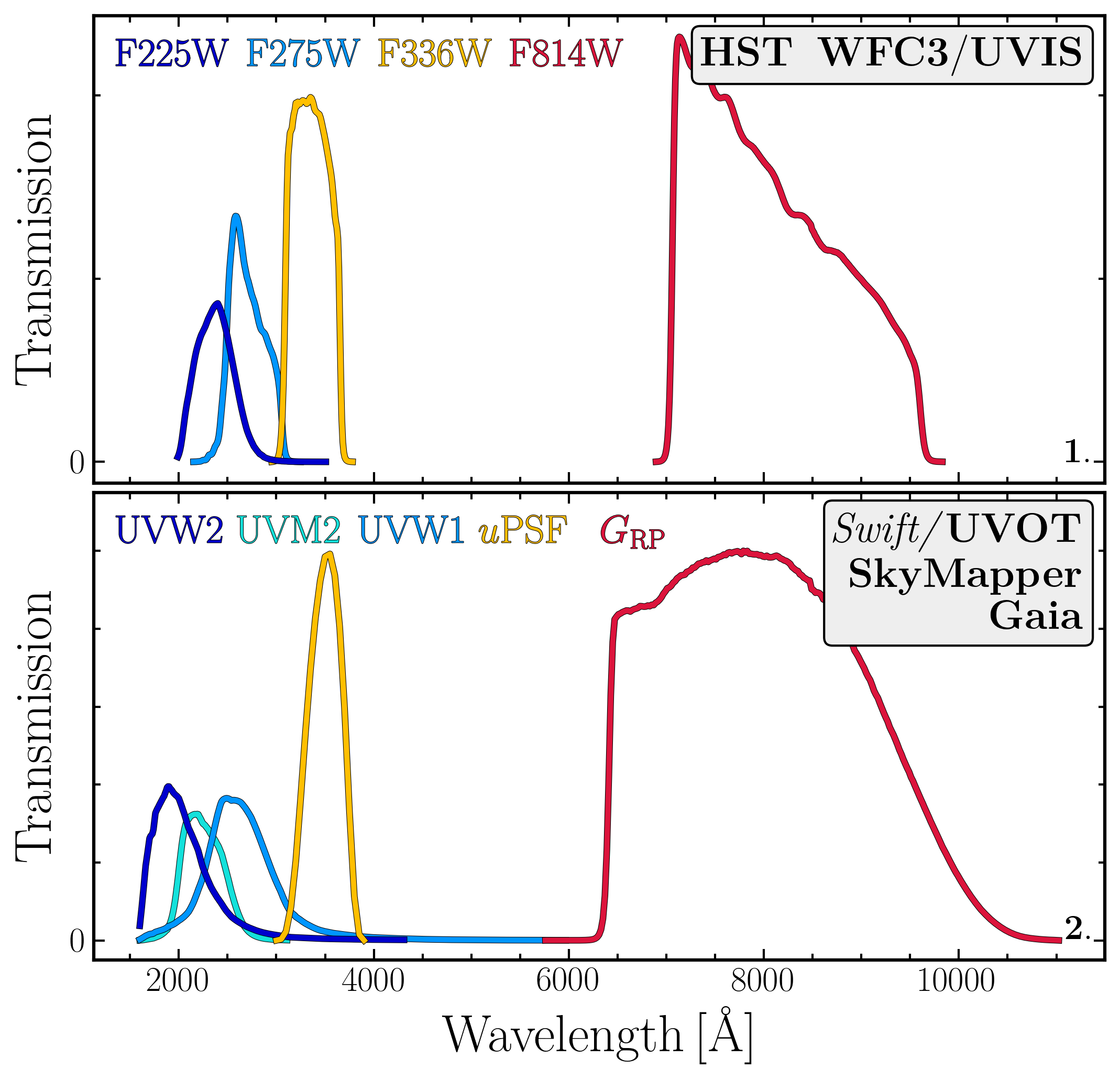}
    \caption{Transmission curves for HST WFC3/UVIS filters (top panel, F225W, F275W, F336W and F814W) and \textit{Swift}/UVOT, SkyMapper and Gaia filters (bottom panel, UVW2, UVM2, UVW1, $u$, $\rp$).}
    \label{fig:filters}
\end{figure}

The transmission curves\footnote{Filters transmission curves are available at \url{http://svo2.cab.intacsic.es/svo/theory/fps3/}} of HST, UVOT, SkyMapper and Gaia filters are shown in Fig.~\ref{fig:filters}, illustrating the correspondence between the HST filters used to study UVdim stars in the MCs \citep[i.e. UVIS F2225W/F275W, F336W and F814W][]{milone2023b}, and the filters used in this study for OCs. Based on the comparison displayed in Fig.~\ref{fig:filters}, we find that the combination of one of \textit{Swift}/UVOT, with SkyMapper $u$ and Gaia $\rp$ provides the best match to HST colours.

\subsection{\textit{Swift} UVOT Stars Survey}
The \textit{Swift}/UVOT Stars Survey includes photometry for 103 OCs in three UV filters: UVW1 ($\lambda_\mathrm{eff} = 2600 \AA$), UVM2 ($\lambda_\mathrm{eff} = 2246 \AA$), and UVW2 ($\lambda_\mathrm{eff} = 1928 \AA$). These filters provide comprehensive coverage of the NUV spectral range, enabling the detection of hot, UV-bright stars. We refer to \citet{siegel2014, siegel2019} for a detailed description of the observing strategy and data-reduction. We only mention here that the photometric magnitudes released in the public catalogues are derived from Point Spread Function (PSF) photometry, obtained combining all images within $7'.5$ from the cluster centres \footnote{This is true for most of the clusters, but there are some examples where the images span a larger field of view.}. Following the prescriptions of \citet{siegel2014, siegel2019}, we excluded stars brighter than the saturation limits of AB magnitudes 12.52, 12.11, and 12.68 in the UVW1, UVM2, and UVW2 filters, respectively. 
Additionally, we selected well measured stars based on the uncertainties provided for each analysed band, and the \texttt{CHI} and \texttt{SHARP} photometric quality parameters. Briefly, \texttt{CHI}  indicates the quality of the PSF fit, with higher values typically corresponding to poor fits or blended sources, while the \texttt{SHARP} parameter measures the intrinsic sharpness of a source: values near zero correspond to point-like sources (e.g. stars), while significant deviations may indicate extended objects, cosmic rays, or blends. 

Cluster members have then been selected by cross-matching UVOT photometric catalogues with cluster members from \citet{hunt2024}, determined by means of Gaia DR3 precise proper motions and parallaxes. We also refer to \citet[][and references therein]{hunt2023} for a detailed description of the methodology. For consistency, all cluster properties, e.g. ages, masses, discussed throughout the article are derived in \citet{hunt2024}. 

To cross-match UVOT photometric catalogues with Gaia data, we transformed Gaia coordinates to the UVOT reference epoch, accounting for proper motions, and retained matches within a maximum separation of 1.5 arcsec. To assess potential mismatches, we examined alternative matches within 5 arcsec by visually evaluating their positions in Gaia-UVOT CMDs. We found that the primary match within 1.5 arcsec was always consistent with the cluster CMD, whereas secondary matches were not. Finally, we visually inspected each CMD to confirm adequate photometric quality, and restricted our final sample to clusters younger than 2\,Gyr.

\subsection{SkyMapper Southern Sky Survey}

SkyMapper DR4 provides high quality $ugriz$ photometry for millions of stars in the southern hemisphere. In this work, we employ the $u$-band PSF photometry ($u_\mathrm{PSF}$) to complement the UV data from UVOT, allowing a comparison with the HST UV-optical two-colours diagram introduced in \citet{milone2023b}.
To ensure precise $u$-band photometry, we used the quality flags described in \citet{onken2024}, including \texttt{Nimaflags} \texttt{u\_flags} and \texttt{u\_ngood}, selecting only well-measured stars. As done for UVOT photometry, we excluded stars with large photometric uncertainties. As done for UVOT photometry, we excluded stars with large photometric uncertainties. 

The final sample includes 35 clusters with ages between $0.02$\,Gyr (NGC\,2571) to $1.89$\,Gyr (NGC\,2627), with accurate photometry in Gaia, \textit{Swift}/UVOT and SkyMapper photometric bands. As a further selection, we exploited Gaia $G$ vs. $G_\mathrm{BP} - G_\mathrm{RP}$ CMD to select only main sequence and turn-off stars. The final UVOT/SkyMapper/Gaia photometric catalogues do not reach the cluster turn-offs in the youngest and closest clusters of our sample due to saturation in the UVOT bands, but they cover nearly the entire main sequence for clusters older than $\sim 0.2\,\mathrm{Gyr}$.

\section{Results}\label{sec:results}
\begin{figure*}
    \centering
    \includegraphics[width=0.8\textwidth]{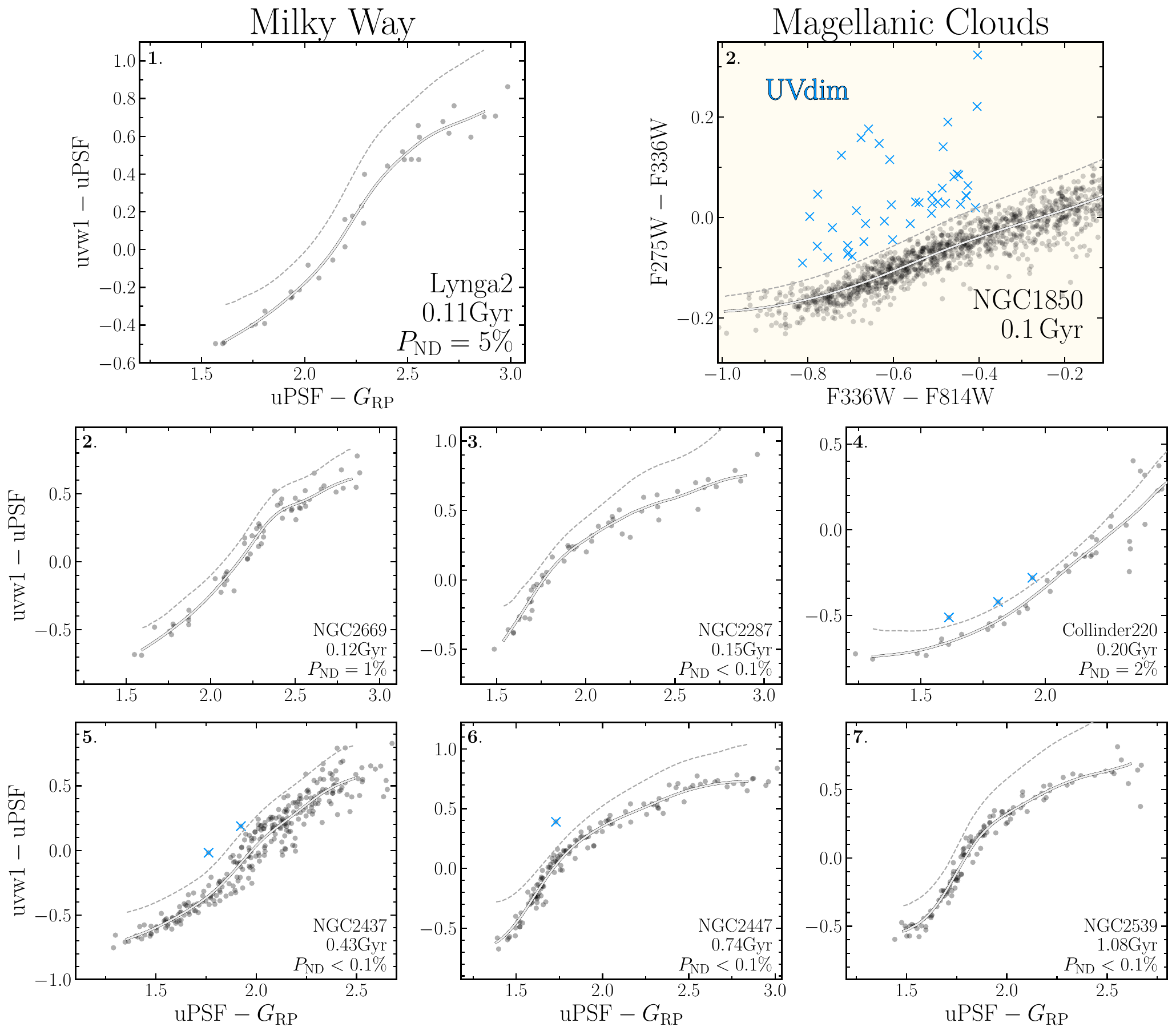}
    \caption{Collection of $\mathrm{UVW1}-\mathrm{uPSF}$ vs. $\mathrm{uPSF}-\rp$ colour-colour diagrams for OCs with ages ranging from 0.1 to 1.1 Gyr. The top-right panel (panel \#2), highlighted with a yellow background, shows the HST diagram of the Large Magellanic Cloud cluster NGC\,1850, used by \citet{milone2023b}. UVdim stars in NGC\,1850 and OCs are marked with azure crosses. Solid lines represent the LOESS fit of the fiducial colour-colour trend, while dashed lines, determined shifting by $3$ times the fiducial LOESS, indicate the threshold for identifying UVdim stars. The bottom right insets show each cluster’s name, age, and the significance of the non-detection (see Sec.\ref{sec:results} for details).}
    \label{fig:mainfig}
\end{figure*}

To identify UVdim stars in young Galactic OCs, we combined photometric magnitudes from \textit{Swift}/UVOT, SkyMapper DR4, and Gaia DR3 to reproduce the HST  F225W/F275W$-$F336W vs. F336W$-$F814W used in \citet{milone2023b} for five young MC clusters. As shown in Fig.~\ref{fig:filters}, UVW1$-u\mathrm{PSF}$ vs. $u\mathrm{PSF} - \rp$ represents the best matching combination and we show in Fig.~\ref{fig:mainfig} displays a collection of UVW1$-u\mathrm{PSF}$ vs. $u\mathrm{PSF} - \rp$ colour-colour diagrams for OCs with ages between 0.1 and 1.1\,Gyr. To facilitate visual comparison, we include the diagram used by \citet{milone2023b} to identify UVdim stars in the Large Magellanic Cloud cluster NGC\,1850 (highlighted in panel 2 with a yellow background). Additional colour-colour diagrams for all clusters and different filter combinations are provided in Appendix~\ref{app:app1}.

In each panel of Fig.~\ref{fig:mainfig}, a dashed grey line shows the threshold used for identifying UVdim stars, which are marked with azure crosses, if present. This threshold is determined by shifting the fiducial line of the colour-colour diagram (shown with solid black-grey lines) by three times its standard deviation. The fiducial line is derived employing the Locally Weighted Regression (LOESS) algorithm described in \citet{cappellari2013}, and its standard deviation is estimated by bootstrapping the sample, repeating the LOESS fit 1,000 times, and taking the $68^\mathrm{th}$ percentile of the resulting distribution. To handle the relatively low number of stars in some of the analysed clusters, we apply a classical wild bootstrap approach. In a nutshell, each data point retains its original 
$x$-value but is assigned a new $y$-value by adding a resampled residual to the fiducial LOESS fit. For a point with residual $r$ (the observed $y$ minus the LOESS prediction), we draw the new $y$-value to be the LOESS fit plus $\pm r$. This procedure retains all original data points (no sampling with replacement) while incorporating the observed residuals into each resampled dataset.

Colour-colour diagrams for all the analysed clusters are presented Appendix~\ref{app:app1}, where UVdim stars are marked again with azure crosses, and clusters with robust identification are highlighted with a light yellow background. More details are provided in Sec.~\ref{sec:discussion}

Figure \ref{fig:uvw1u} (and figures in Appendix~\ref{app:app1}) show that most of the analysed open clusters exhibit a narrow, well-defined trend in the various colour-colour diagrams. Only five clusters, namely NGC\,2301, NGC\,2396, NGC\,2437, NGC\,2658 and NGC\,2447 (shown in panels \# 7, 13, 17, 18 and 22 of Fig.~\ref{fig:uvw1u}-\ref{fig:uvw2uvw1}), have at least one UVdim star candidate in all the colour-colour diagrams. Three clusters, namely Collinder\,220, NGC\,2548 and NGC\,3680, display stars with UVdim-like colours in only one of the analysed colour combinations and are therefore not included in the list of clusters with robust detection. Finally, we find possible UVdim candidates in older clusters (Berkeley\,37, NGC\,2818, NGC\,2508 and NGC\,2627); however, given the lower overall photometric quality and larger colour spread, we do not consider these stars as reliable UVdim candidates. We define as \texttt{robust} the UVdim candidates that are consistently detected across all colour–colour combinations, and as \texttt{possible} those that appear in only a subset of colours. Robust and possible candidates are marked with azure crosses in all panels, while panels corresponding to clusters with robust detections are highlighted with yellow background. Gaia \texttt{source\_id}s for all identified UVdim candidates, along with the colour combinations used to detect them, are provided in Table~\ref{tab:tab1}.

\subsection{Statistical significance of non-detections}
Young Magellanic Clouds clusters are generally more massive than Galactic OCs and therefore host a larger number of stars. Since UVdim stars are expected to account for a small fraction of the total number of stars \citep[see e.g. Fig.~5 of][]{milone2023b}  their detection in Galactic OCs may be hindered by both the lower number of cluster stars and the incompleteness of out photometric catalogues. The combination of these factors reduces the probability of observing UVdim stars in Galactic OCs. To evaluate the statistical significance of the non-detection, we simulated the effects of incompleteness on the expected number of UVdim stars using Magellanic Cloud clusters as benchmark.

For each OC, first created a synthethic cluster with 5\% of stars designated UVdim and a number of stars ($\mathrm{N_{stars}}$) equal to the cluster members in \citet{hunt2024}. The fraction of UVdim is determined based on \citet{milone2023b}, and we repeated the analysis using 2.5\%. Finally, to emulate UVOT photometric catalogues, we accounted for photometric incompleteness by randomly selecting $\mathrm{N_{stars}} \times \mathrm{Completeness}$ stars. Here, completeness has been determined as the ratio between the number of stars in the UVOT/SkyMapper/Gaia catalogues over the total number of stars from \citet{hunt2024}\footnote{For the purposes of this analysis, we assume the Gaia OC data from \citet{hunt2024} to be complete. While this is a simplification, it is a reasonable approximation for the majority of clusters examined in this study. We refer to \citet{hunt2024} for a detailed description of incompleteness}.

We then determined the fraction of UVdim stars found in the completeness-limited sample. For each cluster, we repeated the sampling process 500 times, recording the recovered UVdim fraction in each iteration. To quantify the likelihood of non-detection, we computed the fraction of realisations returning zero UVdim stars, denoted as $\pnd$. By definition, $\pnd = 0$ indicates that at least one UVdim star was detected in all realisations, while $\pnd = 1$ means none were ever detected. A low $\pnd$ suggests that a non-detection is statistically robust; a high $\pnd$ implies that UVdim stars may have been missed due to incompleteness. 

\begin{figure}
    \centering
    \includegraphics[width=0.48\textwidth]{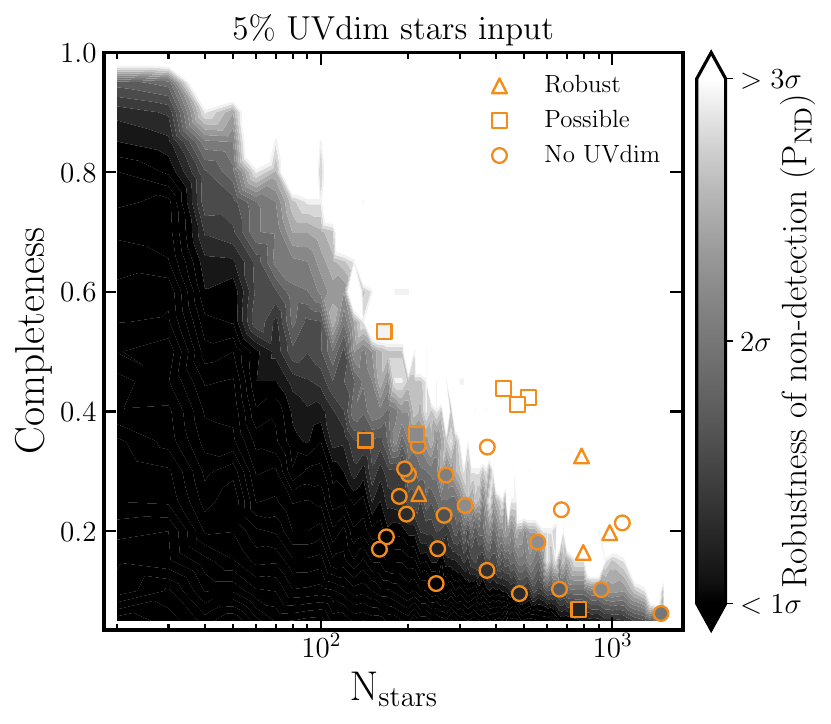}
    \caption{$\pnd$ map as a function of $(N_\mathrm{stars}, \mathrm{Completeness})$ with observed clusters over-plotted. Colour-coding is indicated in the right colorbar, with black colours indicating more robust non-detections. Values of $\pnd$ have been converted into units of standard deviation, assuming a one-tailed probability. Clusters with robust, possible or no UVdim stars detection are marked with triangles, squares and circles, respectively.}
    \label{fig:pmap}
\end{figure}

To explore the dependence of the non-detection significance on number of stars and incompleteness, we repeated the simulations over a grid of parameters. We varied the completeness from 10\% (0.1) to 100\% (1.0) in steps of 0.05 and the number of cluster stars from 10 to 1500 in steps of 10. The resulting $\pnd$ map, displayed in Fig.~\ref{fig:pmap}, shows how the probability of non-detection varies with these parameters. To facilitate the interpretation of $\pnd$, we converted the values of $\pnd$ into units of standard deviation $\sigma$. In the map, white indicates $\pnd$ larger than $3\sigma$, while black colours indicate $\pnd<1\sigma$. Analysed OCs are over-plotted on the map, with the same colour-coding and different markers for robust, possible and non-detection of UVdim stars (as defined in in the previous section). As a further check, we also repeated the steps assuming a lower input fraction of UVdim stars, namely 2.5\%, finding that our conclusions are not significantly affected. It can be seen that, lower number of stars and lower completeness values (bottom left corner of the map in Fig.~\ref{fig:pmap}) yield a lower statistical significance (i.e. higher likelihood of missing UVdim stars). Conversely, larger number of stars and completeness values (top-right corner of Fig.~\ref{fig:pmap}) returns more robust non-detections.

\rowcolors{2}{gray!25}{white}
\begin{table*}
    \centering
    \caption{UVdim stars candidate information. The first part of the table lists UVdim candidate detected in all analysed colour-colour diagrams in Appendix~\ref{app:app1}.}
    \begin{tabular}{ccccccc}
        \hline
        \hline
        \textbf{Cluster} & \textbf{\texttt{source\_id}} & $\mathrm{uvw1}-u_\mathrm{PSF}$ & $\mathrm{uvw2}-u_\mathrm{PSF}$ & $\mathrm{uvw2}-\mathrm{uvw1}$ & $\mathrm{uvm2}-u_\mathrm{PSF}$ & \texttt{robust} \\
        \hline
        NGC\,2301 & 3113578541997114240 & yes & yes & yes & yes & True \\
        NGC\,2396 & 3034562521129150848 & yes & yes & yes & yes & True \\
        NGC\,2437 & 3029160826659377280 & yes & yes & yes & yes & True \\
        NGC\,2447 & 5615576491486822272 & yes & yes & yes & yes & True \\
        NGC\,2658 & 5639313023392485120 & yes & yes & yes & yes & True \\
        \hline
        \hline
        Collinder\,220 & 5255630091479539584 & yes & no & no & no & False \\
        Collinder\,220 & 5351701431777153536 & yes & no & no & no & False \\
        Collinder\,220 & 5351706757536723072 & yes & no & no & no & False \\
        NGC\,2437 & 3029210991877405312 & no & no & yes & no & False \\
        NGC\,2548 & 3064480537453897728 & yes & no & yes & no & False \\
        Berkeley\,37 & 3109968334229971968 & yes & yes & yes & yes & False \\
        Berkeley\,37 & 3109975584134663552 & no & yes & no & no & False \\
        Berkeley\,37 & 3109976718006011392 & no & yes & no & no & False \\
        Berkeley\,37 & 3109979844742190080 & no & yes & yes & yes & False \\
        Berkeley\,37 & 3109980257059027968 & yes & yes & no & no & False \\
        NGC\,2818 & 5623380167892292352 & no & no & yes & no & False \\
        NGC\,2818 & 5623381306063068032 & no & yes & yes & no & False \\
        NGC\,2818 & 5623386631822488832 & no & no & yes & no & False \\
        NGC\,2818 & 5623386734901699840 & yes & yes & yes & no & False \\
        NGC\,3680 & 5382184567010401792 & yes & no & no & no & False \\
        NGC\,2509 & 5714215638128218880 & no & yes & no & no & False \\
        NGC\,2509 & 5714215741207424640 & no & yes & no & yes & False \\
        NGC\,2627 & 5643760513568537984 & no & no & yes & no & False \\
        NGC\,2627 & 5643766457803209728 & no & no & yes & no & False \\
        \hline
        \hline
    \end{tabular}
\label{tab:tab1}
\end{table*}

\section{Discussion and conclusions}\label{sec:discussion}

\begin{figure}
    \centering
    \includegraphics[width=0.45\textwidth]{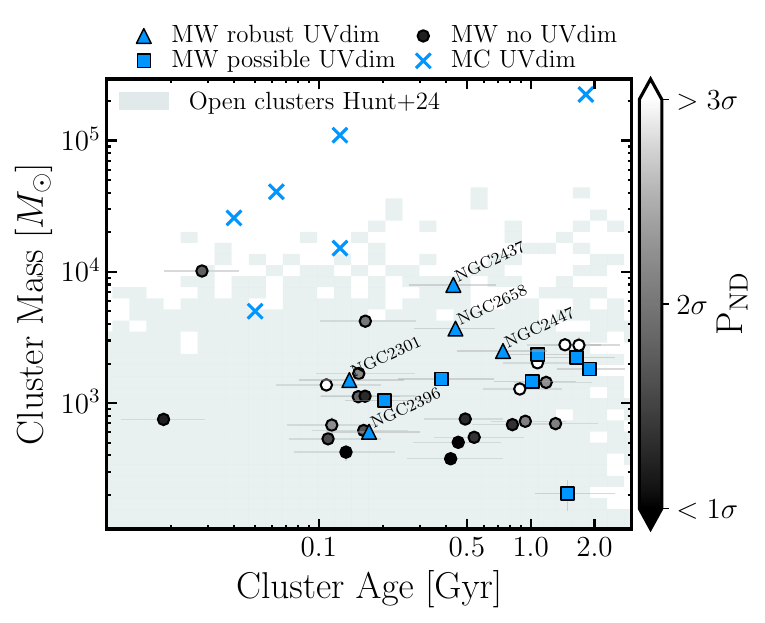}
    \caption{Comparison between Magellanic Clouds and Milky Way clusters ages and masses, separating clusters with and without UVdim, as discussed in Sec.~\ref{sec:results}. Magellanic Clouds are indicated by azure crosses, while Milky Way clusters with robust and possible detection are marked with filled triangles and squares, respectively. OCs without UVdim detections are represented with white-black circles depending on the robustness of the non-detection (right colorbar). We refer to Sec.~\ref{sec:results} for a description of the procedure adopted to compute $\pnd$. The mass-age distribution of the Milky Way clusters from \citet{hunt2024} is shown as grey shaded background.}
    \label{fig:massage}
\end{figure}

We analysed NUV-optical colour-colour diagrams of 35 Galactic OCs to investigate the presence of UVdim stars. 
Out of the 35 analysed OCs, five clusters ($\sim14$\%) show robust UVdim star candidates consistently across all colour–colour diagrams. An additional three clusters ($\sim9$\%) exhibit possible UVdim candidates, but only in one of the diagrams. Another four clusters ($\sim11$\%) show signs of UVdim stars, though the detections are uncertain due to poorer photometric quality of the clusters' catalogues. The remaining 22 clusters ($\sim66$\%) do not show any evidence of UVdim stars. It is worth mentioning that all clusters with robust UVdim stars detections are younger than 1\,Gyr, and out of the 14 OCs younger than 200\,Myr 2 have robust UVdim star detections. Considering the low number of UVdim and cluster stars, and photometric incompleteness, we avoid calculating the fraction of UVdim in the five clusters with robust detections. 

Among the clusters without UVdim detections, 13\% (three clusters) show a non-detection significance of less than $1\sigma$, 65\% (15 clusters) fall in the $1$--$3\sigma$ range, and 22\% (five clusters) exhibit a significance of $3\sigma$ or greater. We also note that the seven clusters with possible UVdim detections, four have $\pnd > 3\sigma$, while the remaining three have lower $\pnd$ values.

Our analysis indicates that UVdim stars are rare or absent in Galactic OCs. This result is in stark contrast with young Magellanic Cloud clusters, where UVdim stars are consistently observed in systems younger than $\sim 200$\,Myr. Indeed, although the potential UVdim-cluster age dependence remains uncertain due to a lack of UV data for clusters between 200\,Myr and 1.5\,Gyr, all clusters younger than 200 Myr with available F225W/F275W photometry exhibit UVdim stars. This suggests that UVdim stars are likely common in young star clusters in the Magellanic Clouds. In contrast, only two out of 14 clusters ($14$\%) younger than 200\,Myr have robust UVdim detections.

We summarise our results in Fig.~\ref{fig:massage}, where we present a comparison between Magellanic Clouds clusters with UVdim (azure crosses) and the MW clusters analysed in this work, differentiating OCs with robust UVdim detections (filled triangles), possible UVdim (filled squares) and without UVdim (white-black circles, colour-coded according to $\pnd$). We refer to the discussion in Sec.~\ref{sec:results} for a description of the three OCs groups. The light-grey background region indicates the locus of the OCs identified in \citet{hunt2024}\footnote{To simplify the plot, we only show where Galactic OCs lie in the cluster mass-age plane, not the density of OCs in this plane.} Figure~\ref{fig:massage} shows that Magellanic Clouds clusters are, on average, at least one order of magnitude more massive than OCs, and the five clusters with robust detections are on the high-mass end of the OCs mass distribution, except for NGC\,2396. We cannot exclude that this is a bias introduced by the sample of analysed clusters, however, we note that it could also suggest a possible mass dependence of the phenomenon. A remarkable exception is NGC\,6649, whose mass is comparable with Magellanic Clouds clusters with similar age, but lack UVdim stars detection.  

The apparent rarity of UVdim stars in young OCs align with the overall lack of split MS in OCs, likely connected to UVdim stars as they mainly populate the blue MS. Indeed, while nearly all MC clusters younger than 700\,Myr show split MSs, only a few OCs, such as NGC\,2287 and NGC\,3532, show a similar feature. Unfortunately, for NGC\,3532 we lack \textit{Swift} UVOT photometry, and NGC\,2287 is heavily affected by saturation; nonetheless, most of its non-saturated stars follow the same trend with no UV-excess. The remaining clusters on the the other hand do not show evidence for a split-MS. While the nature of such difference is beyond the scope of the present work, we find worth mentioning that cluster mass could potentially influence the observed cluster properties. 

Another intriguing difference involves the lifetime of circumstellar disks in the MW and Magellanic Clouds. Recently, \citet{demarchi2024} used JWST NIRSpec spectra to study young stars in the Small Magellanic Cloud cluster NGC\,346 finding that circumstellar disks around sun-like stars in low-metallicity environments survive longer, up to 30\,Myr. This contrasts with the Milky Way, where circumstellar disks typically dissipate within in the first 5/10\,Myr \citep[][and references therein]{komaki2025}. Because stars without disks generally rotate faster than those still bearing disks \citep{venuti2017}, an extended disk lifetime in the Magellanic Clouds could prevent some stars from spinning up via the disk-locking mechanism, thus contributing to create split MSs and bimodal rotation distributions. By contrast, the shorter disks lifetime in the Milky Way would inhibit such rotational bimodality, explaining the absence (or rarity) of split MSs and UVdim in Galactic open clusters.

Finally, \citet{he2024} identified A- and F-type stars with $\halpha$ emission in the young OC NGC\,3532 and proposed that they could be UVdim star candidates with circumstellar decretion disks. However, we also note that UVdim stars identified in \citet{milone2023b} are located on the slow-rotating MS, while UVdim candidates in \citet{he2024} have high rotational velocities. Furthermore, the slow rotating nature of UVdim stars in the 1.5 Gyr-old NGC\,1873 has been recently confirmed by \citet{leanza2025} by means of MUSE spectra.

While the nature of UVdim stars is subject of active debate, we find that they seem to be rarer in Galactic OCs than in Magellanic Clouds clusters. Possible differences may be caused by the different masses, metallicities and/or circumstellar disks lifetimes of Milky Way and Magellanic Clouds clusters. Future works combining photometry and spectroscopy of confirmed UVdim candidates will possibly add important constraints.

\section*{Acknowledgments}
 EPL acknowledges support from the ``Science \& Technology Champion Project'' (202005AB160002) and from the ``Top Team Project'' (202305AT350002), all funded by the ``Yunnan Revitalization Talent Support Program''. This work has received funding from  ``PRIN 2022 2022MMEB9W - \textit{Understanding the formation of globular clusters with their multiple stellar generations}'' (PI Anna F.\,Marino),  and from INAF Research GTO-Grant Normal RSN2-1.05.12.05.10 -  (ref. Anna F. Marino) of the ``Bando INAF per il Finanziamento della Ricerca Fondamentale 2022''.  The national facility capability for SkyMapper has been funded through ARC LIEF grant LE130100104 from the Australian Research Council, awarded to the University of Sydney, the Australian National University, Swinburne University of Technology, the University of Queensland, the University of Western Australia, the University of Melbourne, Curtin University of Technology, Monash University and the Australian Astronomical Observatory. SkyMapper is owned and operated by The Australian National University's Research School of Astronomy and Astrophysics. The survey data were processed and provided by the SkyMapper Team at ANU. The SkyMapper node of the All-Sky Virtual Observatory (ASVO) is hosted at the National Computational Infrastructure (NCI). Development and support of the SkyMapper node of the ASVO has been funded in part by Astronomy Australia Limited (AAL) and the Australian Government through the Commonwealth's Education Investment Fund (EIF) and National Collaborative Research Infrastructure Strategy (NCRIS), particularly the National eResearch Collaboration Tools and Resources (NeCTAR) and the Australian National Data Service Projects (ANDS). This work has made use of data from the European Space Agency (ESA) mission
{\it Gaia} (\url{https://www.cosmos.esa.int/gaia}), processed by the {\it Gaia}
Data Processing and Analysis Consortium (DPAC,
\url{https://www.cosmos.esa.int/web/gaia/dpac/consortium}). Funding for the DPAC
has been provided by national institutions, in particular the institutions
participating in the {\it Gaia} Multilateral Agreement.

\section*{Data Availability}
Relevant data underlying this work is currently publicly available at the following repositories: Gaia Open clusters catalog from \citet{hunt2024}: \url{https://cdsarc.cds.unistra.fr/viz-bin/cat/J/A+A/686/A42}; the Swift UVOT Stars Survey from \citet{siegel2019}: \url{https://cdsarc.cds.unistra.fr/viz-bin/cat/J/AJ/158/35}; and SkyMapper Souther Sky Survey \citep{onken2024}: \url{https://skymapper.anu.edu.au/tap/}. Additional data and figures will be shared upon reasonable request to the corresponding author.

\bibliographystyle{mnras}
\bibliography{main}

\begin{appendix} 
\section{Additional two-colour diagrams}\label{app:app1}
\begin{figure*}
    \centering
    \includegraphics[width=0.9\textwidth, trim={0cm 0cm 0cm 0cm}, clip]{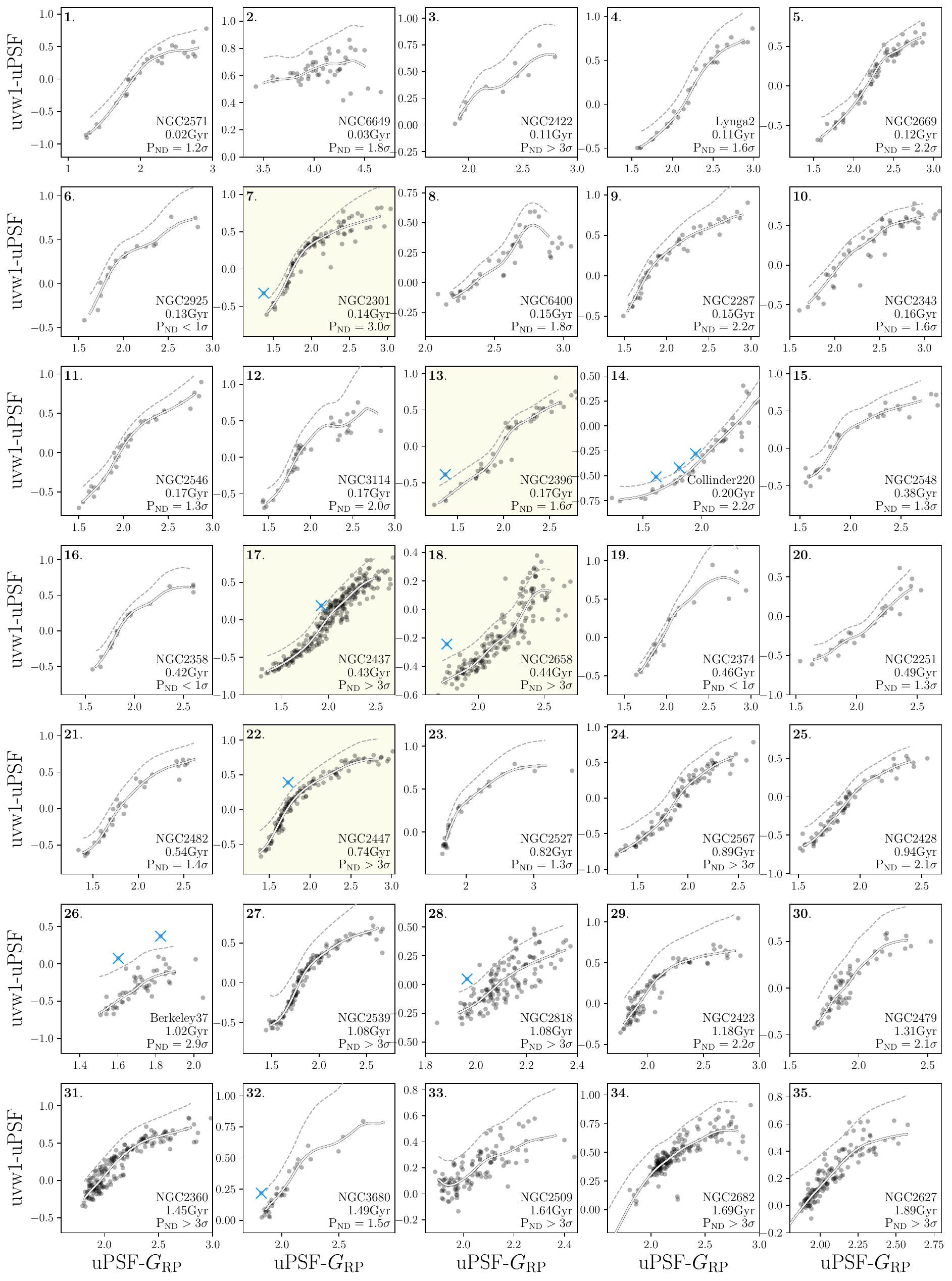}
    \caption{$\mathrm{UVW1} - \mathrm{u_{PSF}}$ vs. $\mathrm{u_{PSF}} - \mathrm{G_{RP}}$ Two-colours diagram for the 35 analysed clusters with both \textit{Swift}/UVOT, SkyMapper DR4 and Gaia DR3 photometry.}
    \label{fig:uvw1u}
\end{figure*}

\begin{figure*}
    \centering
    \includegraphics[width=0.9\textwidth, trim={0cm 0cm 0cm 0cmcm}, clip]{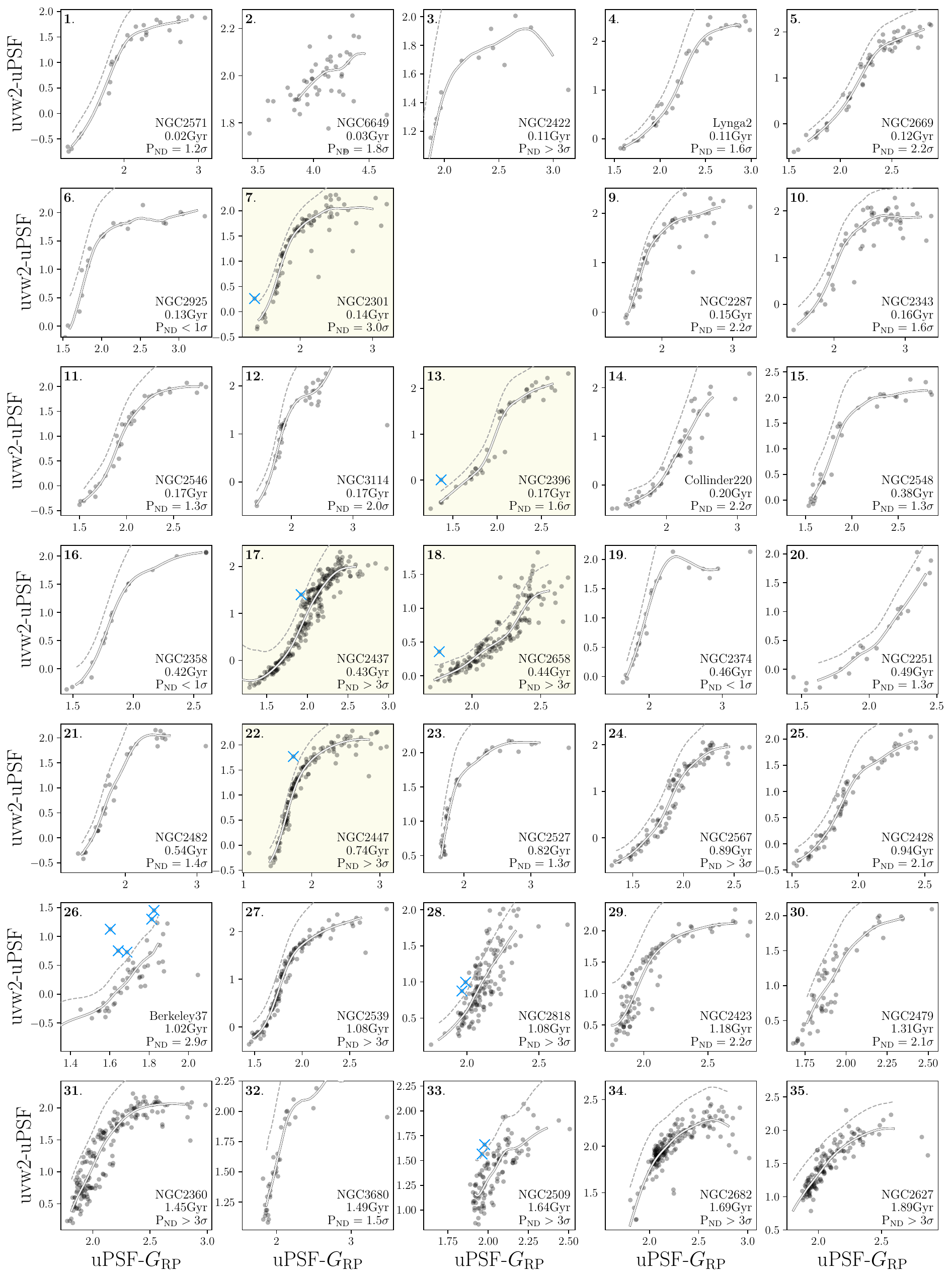}
    \caption{Same as Fig.~\ref{fig:uvw1u} with $\mathrm{UVW2} - \mathrm{u_{PSF}}$ vs. $\mathrm{u_{PSF}} - \mathrm{G_{RP}}$.}
    \label{fig:uvw2u}
\end{figure*}

\begin{figure*}
    \centering
    \includegraphics[width=0.9\textwidth, trim={0cm 0cm 0cm 0cmcm}, clip]{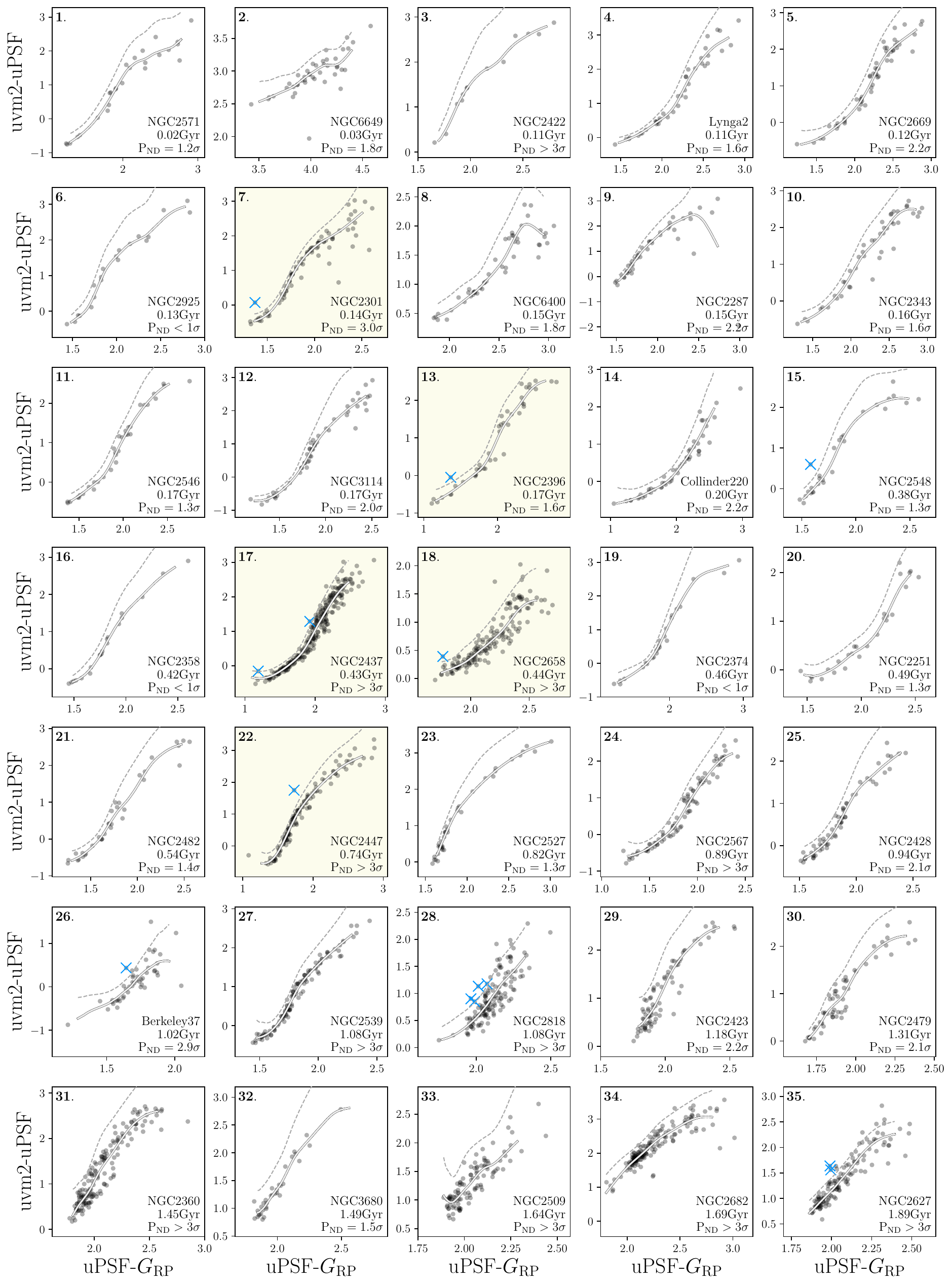}
    \caption{Same as Fig.~\ref{fig:uvw1u} with $\mathrm{UVM2} - \mathrm{u_{PSF}}$ vs. $\mathrm{u_{PSF}} - \mathrm{G_{RP}}$.}
    \label{fig:uvm2u}
\end{figure*}

\begin{figure*}
    \centering
    \includegraphics[width=0.9\textwidth, trim={0cm 0cm 0cm 0cmcm}, clip]{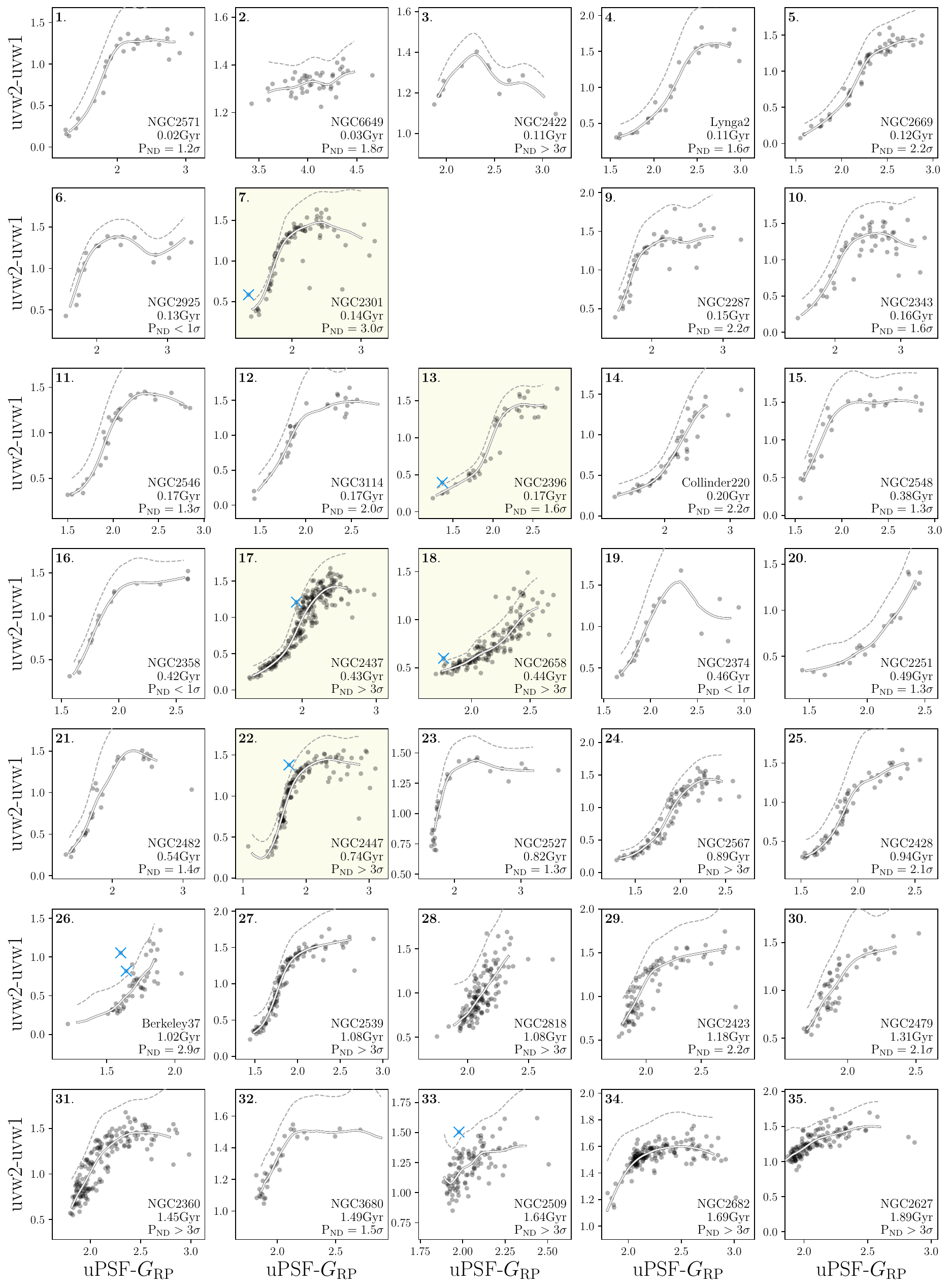}
    \caption{Same as Fig.~\ref{fig:uvw1u} with $\mathrm{UVM2} - \mathrm{UVW1}$ vs. $\mathrm{u_{PSF}} - \mathrm{G_{RP}}$. }
    \label{fig:uvw2uvw1}
\end{figure*}

We present in Figs.~\ref{fig:uvw1u}-\ref{fig:uvw2uvw1} colour-colour diagrams for all 35 analysed clusters and for different colour combinations. As in Fig.\ref{fig:mainfig}, UVdim stars are marked with azure crosses, while clusters with robust detection (i.e. UVdim stars identified in all colour combinations) are highlighted with yellow background. In each panel, the mean trends and the threshold adopted to identify UVdim are displayed with solid and dashed lines, respectively, while cluster name, age and $\pnd$ values are indicated in the bottom-right corner. Empty panel in Fig.~\ref{fig:uvw2u}-\ref{fig:uvw2uvw1} refer to NGC\,6400, for which photometry in UVW2 is missing\footnote{We decided to leave the panels empty, instead of deleting the panel, to facilitate comparison of the same clusters in different figures}. We refer to Sec.~\ref{sec:results} and \ref{sec:discussion} for a desc ription of the procedure adopted to identify UVdim stars, and classify the detection as robust or possible. 
 
\end{appendix}

\end{document}